\documentclass[journal,comsoc]{IEEEtran}
\usepackage{algorithmicx}
\usepackage{algpseudocode}
\usepackage{mdframed}
\usepackage[T1]{fontenc}
\usepackage[numbers,sort&compress]{natbib}
\usepackage{svg}
\usepackage{graphicx} 
\usepackage{amsmath} 
\usepackage[justification=centering]{caption}
\usepackage{mathrsfs}
\usepackage{amsmath}
\usepackage{amsfonts,amssymb}
\usepackage{cases}
\usepackage{booktabs}
\usepackage{subfigure} 
\usepackage{stfloats}
\usepackage[ruled,vlined]{algorithm2e}
\usepackage{url}

\ifCLASSINFOpdf
\else
\fi
\usepackage{amsmath}
\usepackage[cmintegrals]{newtxmath}
\usepackage[numbers,sort&compress]{natbib}
\usepackage{svg}
\usepackage{graphicx} 
\usepackage{amsmath} 
\usepackage[justification=centering]{caption}
\usepackage{mathrsfs}
\usepackage{amsmath}
\usepackage{cases}
\usepackage{booktabs}
\usepackage{subfigure} 
\usepackage{stfloats}
\usepackage{url}
\usepackage{caption}

\begin{document}
%
\title{How to Cache Important Contents for Multi-modal Service in Dynamic Networks: A DRL-based Caching Scheme}
%
%
%

\author{Zhe~Zhang,~\IEEEmembership{Member,~IEEE,}
        Marc~St-Hilaire,~\IEEEmembership{Senior Member,~IEEE,}
        Xin~Wei,~\IEEEmembership{Member,~IEEE,}
        Haiwei~Dong,~\IEEEmembership{Senior Member,~IEEE,}
        Abdulmotaleb~El Saddik,~\IEEEmembership{Fellow,~IEEE}
\thanks{Zhe Zhang and Xin Wei are with the School of Communications and Information Engineering, Nanjing University of Posts and Telecommunications, Nanjing, Jiangsu, China (E-mail: (zhezhang@njupt.edu.cn; xwei@njupt.edu.cn).}
\thanks{Marc St-Hilaire is with the School of Information Technology, Department of Systems and Computer Engineering, Carleton University, Ottawa, ON, K1S 5B6, Canada (E-mail: marc\_st\_hilaire@carleton.ca).}
\thanks{Haiwei Dong is with Ottawa Research Center, Huawei Technologies Canada, Ottawa, ON K2K 3J1, Canada (E-mail: haiwei.dong@ieee.org).}
\thanks{Abdulmotaleb El Saddik is with the Multimedia Communications Research Laboratory (MCRLab), University of Ottawa, Ottawa, ON K1N 6N5, Canada, and also with the Computer Vision Department, Mohamed Bin Zayed University of Artificial Intelligence, Abu Dhabi, United Arab Emirates (E-mail: elsaddik@uottawa.ca).}}

\maketitle

\begin{abstract}
With the continuous evolution of networking technologies, multi-modal services that involve video, audio, and haptic contents are expected to become the dominant multimedia service in the near future. Edge caching is a key technology that can significantly reduce network load and content transmission latency, which is critical for the delivery of multi-modal contents. However, existing caching approaches only rely on a limited number of factors, e.g., popularity, to evaluate their importance for caching, which is inefficient for caching multi-modal contents, especially in dynamic network environments. To overcome this issue, we propose a content importance-based caching scheme which consists of a content importance evaluation model and a caching model. By leveraging dueling double deep Q networks (D3QN) model, the content importance evaluation model can adaptively evaluate contents’ importance in dynamic networks. Based on the evaluated contents’ importance, the caching model can easily cache and evict proper contents to improve caching efficiency. The simulation results show that the proposed content importance-based caching scheme outperforms existing caching schemes in terms of caching hit ratio (at least 15\% higher), reduced network load (up to 22\% reduction), average number of hops (up to 27\% lower), and unsatisfied requests ratio (more than 47\% reduction).
\end{abstract}


%
\IEEEpeerreviewmaketitle

\section{Introduction}
%
%
%
%
	\IEEEPARstart{S}{ince} haptic contents were introduced to the Internet in 2014 \cite{TI}, they have become one of the drivers for technology innovation, and will continue to shape future Internet applications. Haptic contents allow users to interact with digital environments in a more realistic way. They can be integrated with existing audio-visual dominated multimedia applications to form what we call multi-modal applications. This opens up a whole new range of possibilities for Internet applications, from virtual shopping to telesurgery, virtual remote training, advanced automotive, etc. For instance, a pilot in training could experience a more realistic sensation of turbulence or engine failure by integrating video, audio, and haptic content. Hence, how to efficiently transmit multi-modal contents is the foundation to realize those future Internet applications, and it is also believed to play a vital role in the realization of metaverse \cite{roadmap, changle}. 
	
	Satisfying the transmission requirements of multi-modal contents is the key to guaranteeing users' quality of experience (QoE). However, different modalities of contents have significantly different transmission requirements for the networks. For example, video contents require large bandwidth, while haptic contents have extremely low latency transmission requirements (usually less than 1 ms). Table \ref{requirements} shows the general transmission requirements of haptic, video, and audio contents \cite{cross-modal}. Among all the requirements, the ultra-low latency and the incredibly high level of reliability of haptic content are the most challenging to satisfy, since the existing multi-hop transmission mechanism will inevitably induce extra latency, due to queuing, processing, and transmission.
\begin{table}[h]
    \centering
    \caption{Transmission Requirements for Multi-modal Contents}
    \label{requirements}
    \renewcommand{\arraystretch}{1.5}
\begin{tabular}{|c|c|c|c|}
  \hline
  & Video Content & Audio Content & Haptic Content \\
  \hline
  Latency & $\leq $ 150 ms & $\leq $ 350 ms & $\leq $ 60 ms \\
  \hline
  Jitter & $\leq $ 30 ms & $\leq $ 30 ms & $\leq $ 10 ms \\
  \hline
  Data Loss Rate  & ~30\% & ~20\% & $10^{-5}$\% \\
  \hline
  Data Rate  & [8-200] Mbps & [64-6000] kbps & [100-1000] kbps \\
  \hline
\end{tabular}
\end{table}
	
	The currently being deployed 5G, with features like ultra-reliable low latency communications (URLLC), can make 1 ms latency transmission possible. However, this low latency transmission is only applicable to the network edge, which means the latency that is caused by the multi-hop forwarding in the backbone networks is still higher than 1 ms. Therefore, the instinctive idea to reduce the long distance transmission latency is to bring contents and services closer to the end users. Consequently, edge caching is believed to be an ideal technology to realize that instinct idea \cite{multi-modal-1, multi-modal-2}.

	
	
	Edge caching aims to store frequently requested contents at the edge of networks, e.g., base stations (BSs), edge routers, etc., to reduce transmission latency and traffic load that are caused by redundant data transmissions from remote servers to end users. In general, caching schemes include a caching decision policy and a replacement policy. The caching decision policy decides what content should be cached, while the replacement policy decides what content should be evicted when the cache storage is full. Both the caching decision policy and the replacement policy have impacts on the efficiency of edge caching. Thus, how to design an efficient caching scheme (including the caching decision policy and the replacement policy) is the most important task in edge caching. 
	
	Traditional edge caching schemes usually consider content popularity as the key indicator to choose the important content for caching \cite{tradi-2, tradi-3}. Although some of them also consider other factors, e.g., topology \cite{RPC, ezABC, interaction}, user mobility \cite{mobility-1, mobility-2, mobility-3}, etc., to make caching decisions, they assume the impacts that are caused by all these factors are static, and assigning fixed weights to these factors to indicate their impacts. Thus, these caching schemes cannot be applied to dynamic networks.
 
 In recent years, artificial intelligence (AI) models, especially reinforcement learning (RL) or deep reinforcement learning (DRL) have become popular for further improving caching efficiency, since the caching problem can be modeled as a Markov decision process (MDP) which is extremely suitable for using DRL models. However, most existing edge caching schemes are limited to single-modal content types, such as video contents \cite{videoCaching-1, videoCaching-2, videoCaching-3} or haptic contents \cite{hapticCaching-1, hapticCaching-2}, and do not provide clear guidelines for caching multi-modal contents. Moreover, the existing DRL-based caching schemes \cite{DRL-1, DRL-2, DRL-3} often neglect the impacts of the network environments. Once the network environment changes, e.g., there are newly released contents, the change of user request pattern, the varying available link bandwidth, etc., the previously trained model becomes invalid (or less effective) for this new network environment.
 

	
	To guarantee users' immersive experience with multi-modal services, edge caching is essential to accurately store "key contents" within dynamic network environments. In this paper, we consider content popularity, content size, content modal type, and network environment to determine the "key contents". Content popularity is a critical factor that directly impacts content importance, and it has been widely considered in existing caching approaches \cite{RPC, ABC, tradi-2, tradi-3, DPWCS, semantic}. Since the caching nodes have limited caching storage, they cannot cache all contents. Consequently, content size should be considered as another determinant when assessing content importance, especially when there are not too many differences among other factors. Furthermore, the content modal type also requires consideration in determining content importance due to varying transmission requirements for different content types. For instance, content such as haptic feedback may demand low latency, while video content may require significant bandwidth. Thus, considering content modal type in evaluating content importance provides better network resource utilization. Additionally, the notion of "key contents" could change depending on the network environment. For example, in a remote virtual training for professions scenario, the VR (virtual reality)/AR (augmented reality) contents could be more important to cache if the trainee's surrounding scenes need to be rendered, while the haptic feedback would be more important if the trainee is touching some objects. Hence, the network environment also plays a vital role in deciding contents' importance. Therefore, how to select the key content to cache is challenging for caching multi-modal contents in dynamic network environments.

	Motivated by this, we first leverage dueling double deep Q networks (D3QN) to adaptively evaluate the content importance in dynamic network environments. By defining the current network environment as the state set, and defining the evaluations of content importance as the action set, the proposed approach can overcome the impact of dynamic network environments. Based on the evaluated content importance, we then propose a content importance-based caching scheme that consists of a caching decision policy and a replacement policy. Consequently, the proposed caching scheme has the ability to automatically cache the most important contents for multi-modal services, and evict the less important contents when the cache storage is full in dynamic network environments.
	
	The main contributions of this article are summarized as follows:
	\begin{itemize}
	\item Unlike existing caching schemes that primarily rely on content popularity to select important contents for caching, we consider more factors, such as content modal type, content popularity, content size, dynamic network environments (including the change of available link bandwidth, variations of user request pattern, shifts in current number of requests, and the introduction of new content), etc., to efficiently cache important contents in dynamic networks. To the best of our knowledge, this is the first comprehensive caching approach that can cache multi-modal contents in dynamic networks.
    \item To enable the proposed content importance-based caching scheme to adaptively select important contents for caching in dynamic networks, we propose a content importance evaluation model from an automatic control perspective by leveraging the D3QN model. In this way, the proposed content importance evaluation model can automatically evaluate each modal content's importance in dynamic networks.
    \item Based on the contents' evaluated importance value, we propose a content importance-based caching decision policy and a content importance-based replacement policy that are both light-weight and efficient for caching multi-modal contents in dynamic networks.
    \item In addition to traditional evaluation metrics, such as hit ratio, average number of hops, and reduced network load, we also introduce unsatisfied requests ratio to evaluate the efficiency of the proposed caching scheme, since this metric can reflect users' QoE. Simulation results demonstrate that the proposed caching scheme outperforms existing schemes in terms of all the above mentioned metrics, e.g., reducing the average number of hops by up to 30\%, increasing the hit ratio by up to 7\%, reducing the network load by more than 13\%, and improving the reduction of unsatisfied requests ratio by up to 10\%.
	\end{itemize}
	
	The remainder of this paper is organized as follows: The related research work is illustrated in Section \uppercase\expandafter{\romannumeral2}. We present the system model in Section \uppercase\expandafter{\romannumeral3}. In Section \uppercase\expandafter{\romannumeral4}, we elaborate the details of the proposed DRL-based caching scheme which includes a content importance evaluation model and a content importance-based caching model. The simulation settings and results are shown in Section \uppercase\expandafter{\romannumeral5}. In Section \uppercase\expandafter{\romannumeral6}, we conclude this paper and outline our future research directions.


\section{Related Work}
In this section, we present the related work about caching schemes for video, haptic, multi-modal contents, and more recently proposed AI-based caching schemes.

\subsection{Video Content Caching Schemes}
Compared to other Internet contents, such as audio and text, video contents including high definition (HD) videos, AR, and VR contents account for the majority of the Internet traffic. Caching them at the network edge can significantly reduce the backbone network traffic and the transmission latency. To find suitable video contents to cache, popularity is believed as one of the important factors for video caching \cite{videoCaching-1}, and has been widely considered in existing works. For instance, \cite{RPC} and \cite{ABC} cache popular videos by assigning a caching threshold at each caching node. The difference between them is that \cite{RPC} considers routers' topology position, while \cite{ABC} considers content's age to make caching decisions. Other factors, such as user preference \cite{access}, CPU, RAM, and disk resources \cite{S-Cache} can also be considered for designing caching scheme.

Zhao \emph{et al}. proposed a popularity-based and version-aware caching scheme (PVCS) \cite{videoCaching-1} for multi-version video-on-demand (VoD) systems. The proposed approach can obtain the maximum caching benefit by caching videos that have the highest popularity and the smallest file size. The replacement policy dynamically evicts videos, leveraging transcoding relations among different versions to enhance version-aware caching profit.

In recent years, leveraging AI models to improve caching efficiency has become the trend since they can make more accurate predictions and take intelligent decisions. In a recent study \cite{prediction}, long short term memory (LSTM) is utilized to predict videos' popularity for proactive caching in mobile edge computing (MEC) scenarios. More specifically, the proposed two-step model firstly predicts the most prevalent movie genre in different time slots of the day, then predicts the movies' future views in different time slots of the day. Based on the predicted popularity, the proposed caching scheme can proactively cache videos at edge servers to reduce backhaul congestion and access delay. However, the proposed caching scheme requires detailed information about movies to predict their popularity, and it is extremely hard for the network operator to obtain that information. Consequently, the proposed approach cannot be easily implemented in real networks.

Although using AI models can improve caching efficiency, such black-box approaches lack explainability and interpretability, and usually have a huge number of parameters. To overcome this issue, Zhang \emph{et al}. in \cite{optimizing} proposed a white-box approach to cache videos at the network edge. A mathematical model (called HRS model) which is the combination of multiple point processes is proposed to capture the evolution of video popularity. All of its model parameters can be automatically learned by maximizing the Log-likelihood function which is constructed by historical video requests. 

\subsection{Haptic Content Caching Schemes}
By caching haptic feedback data of repeated interactions at the edge of networks, the haptic feedback transmission latency can be significantly reduced \cite{design}. A hybrid caching scheme which includes local caching, device-to-device (D2D) caching, small base station (SBS) caching, and macro base station (MBS) caching is proposed in \cite{hybrid} to reduce the haptic data transmission latency. To fit the proposed hybrid caching scheme, a double segmented least recently used (LRU) cache replacement is designed. The proposed replacement policy divides the caching space into two prioritized parts. If there is a hit, the data will be cached at the low priority part first, and it will be moved to the high priority part if another hit happens. If there are more hits, the cached data will be moved to the queue's tail. In this way, popular data will not be easily evicted, while unpopular data will be replaced soon.

Wang \emph{et al}. in \cite{mobility-3} proposed a proactive caching with delay guarantee (PCDG) scheme for haptic contents in mobile networks. They first modeled the network links using the M/M/1 queuing model, and modeled the caching node using M/M/c queues. Then, the caching problem can be formulated as a mixed integer linear programming (MILP) problem. Since solving the formulated MILP problem is time consuming, a greedy caching algorithm (GRC) which tries to cache haptic content to the nearest node based on the maximum user equipment's moving probability is proposed. 

User devices are leveraged to act as helpers for caching haptic content in \cite{hapticCaching-1}. Since helpers are energy constrained, they are not willing to cache other users' content if there is no incentive. Thus, blockchain technology is adopted to propose an incentive mechanism in which helpers act as consensus users of blockchain and can obtain the computational resource from BSs as the incentive.  

\subsection{Multi-modal Content Caching Schemes}
How to cache the suitable modal of contents plays a vital role in multi-modal communications. The study \cite{AI-enabled} utilized mobile edge servers (MEC) to cache audio-visual contents and haptic contents to enhance the users' immersive experience. The proposed caching mechanism is integrated with communications and computing by designing a joint framework that includes transmission mode selection, caching resource management, and computation overhead control. By using RL, the proposed framework can automatically choose the suitable action to manipulate the corresponding part of the framework to fit the dynamic network environments. However, this work focuses on the conceptual aspects of caching and does not propose a practical caching scheme.

A hybrid caching scheme is proposed in \cite{cross-modal, multi-modal-1} for multi-modal communications. It consists of two parts, the first part is passive caching which can passively cache haptic content by implementing a scale haptic coding method that is similar to the traditional scalable video caching. The second part is active caching which adopts a dynamic least mean square (LMS) prediction model to adjust the storage priority for haptic content caching.

\subsection{AI-based Caching Schemes}
Transfer learning is leveraged in \cite{edge-intelligence} to overcome data sparsity issues in estimating content popularity. By implementing transfer learning, more meaningful information can be provided from the rich data domain to help improve the prediction accuracy of the popularity of the target domain's content, e.g., haptic content.  

Wang \emph{et al}. in \cite{semantic} proposed a content popularity-based deep Q-networks (CP-DQN) caching scheme for multi-modal semantic communications. The caching problem is modeled as a finite MDP problem, and by employing DQN, the proposed caching scheme can make optimal caching decisions for each user based on the content popularity and the storage size of edge nodes. Zhang \emph{et al}. presented a similar approach by formulating the caching replacement problem as a MDP from the optimization perspective in \cite{DDQN}. Compared to \cite{semantic}, a federated deep reinforcement learning (FDRL) framework is proposed to find the optimal caching policy in fog radio access networks (F-RANs). Dueling deep Q-network (DDQN) is used as the DRL model, and horizontal federated learning (HFL) is used as the federated learning technique to solve the resource over-consumption issues during the training and data transmission phases. 

Another distributed DQN-based caching scheme for F-RANs is proposed in \cite{access}. Similarly, the caching problem is also modeled as an optimization problem. The difference between \cite{DDQN} and \cite{access} is that the user preference is predicted and used to predict content popularity. In this way, the proposed approach can address the variations of content popularity to obtain a better hit ratio. 


\subsection{Shortcomings of Existing Works}
There are plenty of excellent works that have been conducted in video content caching, haptic content caching, and multi-modal content caching. However, most of them evaluate the importance of contents to make caching decisions only based on content popularity. Other factors, such as content modal type, user request arrival rate, available link bandwidth, etc., also have strong impacts on the caching efficiency. Moreover, these factors are not static, and their impacts are also varied over time in dynamic networks. For example, if the link is congested, caching more popular contents, e.g., video contents can significantly reduce the traffic load to relieve the congestion; while if the link bandwidth is only enough for meeting the latency requirements of certain types of contents, e.g., videos, caching less popular but critical content, e.g., haptic content can significantly improve users’ QoE compared to caching popular video contents since haptic contents have more strict latency requirements. Therefore, evaluating the importance of contents only by popularity is not a good practice for caching multi-modal contents in dynamic networks. Unfortunately, the effect of a dynamic network environment on the efficiency of multi-modal content caching is still unexplored.

Moreover, there could be newly released contents in dynamic networks. Since existing DRL-based caching schemes, e.g., \cite{DRL-1, DRL-2, DRL-3, semantic, DDQN, access}, that define the elements in the action set as caching what data packets at which node, the newly released contents are not covered in the training phase, which makes existing DRL-based caching schemes cannot handle this situation. To overcome this issue, A DQN-based caching approach (named iCache) is proposed in \cite{iCache} to cache IoT data in dynamic networks. By defining the elements in the action set as the nodes’ caching probability that the DQN agent can adjust, iCache can automatically adjust nodes’ caching probability to fit the dynamic network environments. However, it only works for IoT networks since it only considers data popularity and network environments.



\section{System Model}
In this section, we present the details of mathematical modeling for the proposed caching scheme.

\subsection{Network Model}
\begin{figure}[!htp]
\centering{\includegraphics[width=0.5\textwidth]{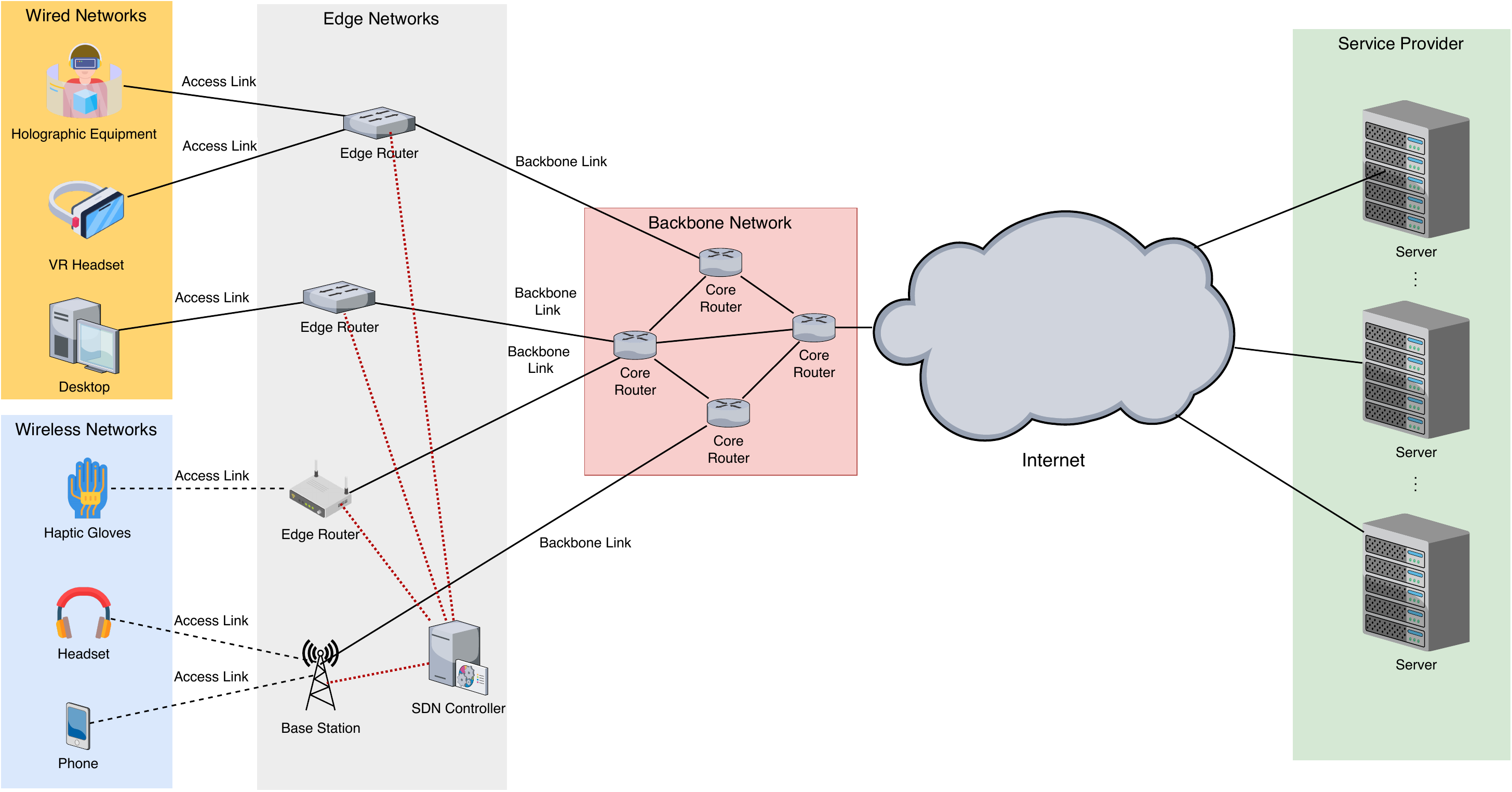}}
\caption{Network model (end users retrieve multi-modal contents from multi-modal service providers through the networks, an SDN controller is implemented in the edge networks to manage and monitor the networks)} 
\label{network-model} 
\end{figure}

In this paper, we consider a multi-modal content (audio, video, and haptic contents) transmission scenario in which users request multi-modal contents from remote content provider servers over wireless (e.g., 5G) or wired networks. The assumed scenario is illustrated in Fig. \ref{network-model}. 

As shown in Fig. \ref{network-model}, edge nodes (could be BSs for wireless links, and edge routers for wired links) are connected to users via wireless networks or wired networks, and are connected to the backbone network through optical links. We use $E=\{e_1, e_2, ..., e_n, ..., e_N\}$ to indicate the set of edge nodes, and $N$ is the total number of edge nodes. Any given edge node $e_n$ can be represented by $e_n = \{al_{bw}(e_n), bl_{bw}(e_n), c_{size}(e_n)\}$, where $al_{bw}$ is the access link (between end users and edge nodes) bandwidth, $bl_{bw}$ is the backbone link (between edge nodes and backbone network) bandwidth, and $c_{size}$ is the cache size. We assume that multi-modal content providers provide either one-modal content or combinations of multi-modal contents to the end users. We denote $D=\{d_1, d_2, ..., d_m, ..., d_M\}$ as the set of all provided multi-modal contents, where $M$ is the total number of all provided multi-modal contents, $D_c(e_n)$ is the set of cached contents at edge node $e_n$. Moreover, each content can be further represented by $d_m=\{o(d_m), r_{bw}(d_m), r_l(d_m), s(d_m)\}$, where $o(d_m)$ is the modality of the content, $r_{bw}$ is the bandwidth requirement, $r_l(d_m)$ is the latency requirement, and $s(d_m)$ is the total size of $d_m$. To meet content $d_m$'s latency requirement transmitting from node $e_n$, it should follow the following constraint:
\begin{align}
\frac{al_{bw}(d_m, e_n)}{s(d_m)} &\leq r_l(d_m), \\
al_{bw}(d_m, e_n) &\leq al_{bw}(e_n),
\end{align}
where $al_{bw}(d_m, e_n)$ stands for the available link bandwidth that can be assigned to content $d_m$ from node $e_n$.

\subsection{User Request Model}
Since the user behaviors for requesting multi-modal contents are strongly affected by users' daily routines, the user request arrival rate varies over time. Based on the change of user request arrival rate, the network can be divided into idle periods and peak periods in general. As a corollary, content popularity is more important for making caching decisions to reduce network load in the peak periods; while its importance could be reduced for making caching decisions in the idle periods as improving user QoE is more vital.

To describe how users' request arrival rate changes over time, we denote all the requests for multi-modal contents from edge node $e_n$ at time $t$ by $Req(e_n, t) = \{req_{d1}(e_n, t), req_{d2}(e_n, t), ..., req_{dm}(e_n, t), ..., req_{dM}(e_n, t)\}$, where $req_{dm}(e_n, t)$ represents the total number of requests for content $d_m$ from node $e_n$ at time $t$, which can represent content $d_m$'s local popularity at time $t$. Therefore, the total number of requests $Req$ that are generated in the network can be calculated as follows:
\begin{equation}
\begin{split}
Req =& \sum_{e_n \in E}\sum_{t \in T}{Req(e_n, t)} \\
    =& \sum_{e_n \in E}\sum_{m \in M}\sum_{t \in T}{req_{d_m}(e_n, t)}.
\end{split}
\end{equation}


\subsection{Content Importance Model}
Finding proper contents to cache plays a vital role in reducing content transmission latency and improving user QoE. However, the importance of contents varies over time. To investigate how content importance varies in dynamic networks, we use $I(d_m, e_n, t)$ to indicate the importance of content $d_m$ for edge node $e_n$ at time $t$. Notably, the importance of content $d_m$ for different BS could be different, and it can be calculated as follows:
\begin{equation}
\begin{split}
I(d_m, e_n, t) =& \omega_1a(e_n, t)+\omega_2\frac{req_{dm}(e_n, t)}{Req(e_n, t)} \\
                &+ \omega_3 o(d_m)+\omega_4 s(d_m)
\end{split}
\end{equation}
where $\omega_1, \omega_2, \omega_3, \omega_4$ are configurable weights, $a(e_n, t)$ stands for the available access link bandwidth of $e_n$ for transmitting the required content, $ Req(e_n, t)$ represents the total number of requests in $e_n$ at time $t$, $req_{dm}(e_n, t)/Req(e_n, t)$ indicates the proportion of requests for content $d_m$' out of the total number of requests $Req(e_n, t)$, $o(d_m)$ is the modality of content $d_m$, $s(d_m)$ is the size of content $d_m$. By adjusting the above configurable weights, we can dynamically evaluate the content importance in any given edge node and at any time.

\subsection{Caching Model}
Without caching, the multi-modal contents have to go through the entire Internet to user devices, which inevitably induces large latencies that are unacceptable for haptic contents. To reduce the large latency that is caused by the long distance transmission, we implement caching at the edge of the networks, e.g., BSs and edge routers. 

The caching model includes a caching decision policy (denoted as $\Phi$) and a replacement policy (denoted as $\Lambda$). The caching decision policy decides what content should be cached locally, while the replacement policy decides what content should be evicted first when the cache memory is full. For edge node $e_n$, its caching state can be described as $C_{e_n}(t) = \{c_{e_n}(d_1, t), c_{e_n}(d_2, t), ..., c_{e_n}(d_m, t), ..., c_{e_n}(d_M, t)\}$, where $c_{e_n}(d_m, t) = 0$ if content $d_m$ is not cached at edge node $e_n$; while $c_{e_n}(d_m, t) = 1$ if content $d_m$ is cached at edge node $e_n$. Since the caching space of each caching node is limited, we have the following condition to express that the total size of the cached contents at node $e_n$ cannot exceed its caching size:
\begin{equation}
\sum_{d_m \in D_c(e_n)}{s(d_m)} \leq c_{size}(e_n).
\end{equation}

\section{Proposed DRL-based Caching Scheme}

The existing DRL-based caching approaches usually define the action set in a straightforward way (i.e., deciding what content should be cached at which node. Thus, the caching problem is turned into an optimization problem under the given state set). However, once the state set changes (i.e., the user request patterns change, or there are newly released contents), the previously trained model has to be re-trained according to the new state to obtain optimal solutions. Therefore, those approaches can only achieve the optimum for static network environments. Moreover, the agent in existing DRL-based approaches has to make caching decisions for each content, which will inevitably introduce extra latency. However, the haptic contents have stringent transmission latency requirement, the extra latency that is caused by the agent could have significant impacts on transmitting haptic contents. 

To tackle the above mentioned shortcomings, the proposed DRL-based caching scheme separates the caching decision making process into two sub-processes: the content importance evaluation process and the caching decision making process. The content importance evaluation process can be handled by the content importance evaluation model by leveraging the D3QN model. The latency that is caused by performing D3QN is acceptable for evaluating contents' importance, since contents' importance will not change dramatically within a short time period. Based on the contents' evaluated importance, the caching model can easily make placement and replacement decisions. Since the computational complexity of the proposed caching schemes is extremely low, its processing latency can be ignored. Thus, it would not affect the transmission of haptic contents.

\begin{figure}[!htp]
\centering {\includegraphics[width=0.5\textwidth]{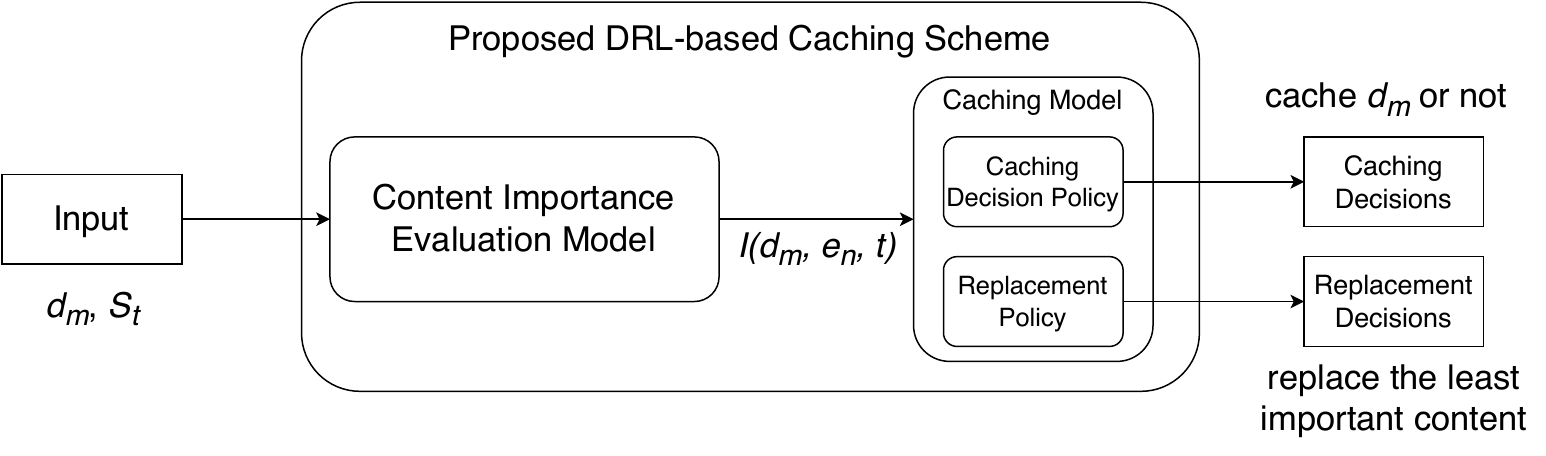}}
\caption{Framework of the proposed DRL-based caching scheme} 
\label{architecture} 
\end{figure}

The framework of the proposed DRL-based caching scheme is depicted in Fig. \ref{architecture}. In this section, we first explain the implementation details of the content importance evaluation model and then, we illustrate how the content importance-based caching model works by introducing its caching decision policy and replacement policy.

\subsection{Content Importance Evaluation Model}
As described in Section \uppercase\expandafter{\romannumeral3}. C, the key objective of the content importance evaluation model is to dynamically evaluate the content importance based on the current network environments. To achieve this objective, we first model the above evaluation process as an MDP from an automatic control perspective rather than from an optimization perspective. Then, we leverage the D3QN model to dynamically evaluate the importance of a given content based on the current network environments.

 \subsubsection{Problem Formulation}

Since the objective of the content importance evaluation model is to evaluate the importance of any given content based on the current network environments, it can be modeled as an MDP problem. 

More specifically, we define the state set as $\mathcal{S}$, then the state at time $t$ can be denoted as $\mathcal{S}_t \in \mathcal{S}$, where $\mathcal{S}_t = \{a(e_nt, t), req_{d_m}(e_n, t),Req(b_n, t), o(d_m), s(d_m)\}$ indicating the information that the agent can observe, including available link bandwidth, content $d_m$'s local popularity that is represented by the number of requests for $d_m$ from node $e_n$ at time $t$, the total number of requests, and the modal type of $d_m$.

We define the action set, denoted as $\mathcal{A}$, as the set of possible evaluation values (integers) and at any given time $t$, we have $\mathcal{A}_t \in \mathcal{A}$. By finding the optimal parameters $\theta$ of the deep neural network of D3QN, the content's importance at state $\mathcal{S}_t$ can be estimated. Consequently, the agent can dynamically evaluate the importance of a given content (selecting one value from the action set) based on the current network environment. In this way, the caching problem is modeled from the automatic control perspective.

$\mathcal{R}$ is designated as the reward set, and we use $\mathcal{R}_t$ to demonstrate the direct reward after taking cation $\mathcal{A}_t$ in state $\mathcal{S}_t$, where $\mathcal{R}_t \in \mathcal{R}$. By maximizing the reward, the importance of any given content can be accurately evaluated.


 \subsubsection{D3QN Model}
 In practical scenarios (real networks), the agent may not have complete knowledge of all elements of the modeled MDP. This can lead to uncertainty in how the environment will respond to the agent's actions, since the transition probabilities are unknown prior. Therefore, the agent cannot accurately predict the results of its actions and has to rely on trial and error to learn how to make optimal decisions. Consequently, the model-free DRL models will be the suitable technique to evaluate the content importance in real network scenarios.

 Deep Q-network (DQN) is a basic model-free DRL model and its effectiveness has been proven in the gaming domain \cite{deepmind, game2}. By replacing the Q-table in Q-learning with a function approximator, e.g., a deep neural network, DQN can overcome the dimensionality curse of Q-learning. The estimated Q-values of taking action $\mathcal{A}_t$ in state $\mathcal{S}_t$, $V(\mathcal{S}_t)$ is the state value of state $\mathcal{S}_t$ is denoted as $Q(\mathcal{S}_t,\mathcal{A})$, and it can be calculated by minimizing the following loss:
 \begin{equation}
 \begin{split}
L(\theta)=&[ (\mathcal{R}_t + \gamma max_{\mathcal{A}_{t+1}}Q(\mathcal{S}_{t+1}, \mathcal{A}_{t+1}; \theta^{target})) - \\
&Q(\mathcal{S}_t,\mathcal{A};\theta) ]^2
\end{split}
\end{equation}
where $R_t$ is the immediate reward at time $t$, $\gamma$ is a discount factor, $\mathcal{A}_{t+1}$ is the action that the agent chooses at time $t+1$, $\theta$ stands for the hyperparameters of the deep neural network, and $\theta^{target}$ is denoted as the hyperparameters of the target deep neural network.  $(R_t + \gamma max_{\mathcal{A}_{t+1}}Q(\mathcal{S}_{t+1}, \mathcal{A}_{t+1}; \theta^{target}))$ is the temporal difference (TD) target, while $Q(\mathcal{S}_t,\mathcal{A};\theta)$ is the TD error. Through measuring the difference between the TD target and the TD error, $\theta$ can be optimized. 

However, its effectiveness could be significantly degraded in real-world problems due to the over-estimation of Q-values, i.e., $Q(\mathcal{S}_t,\mathcal{A})$. To overcome the over-estimation issues of DQN and improve the stability, multiple models are proposed. For example, double DQN (DDQN) is proposed by maintaining two separate networks to reduce the over-estimation: one to determine the action selected by greedy policy, while the other one to determine its Q value. Dueling DQN is proposed by utilizing a dueling architecture to separate the Q-value estimation for each state into two separate parts: estimations for the value of the state itself, and estimations for the advantage of each possible action in that state. Since DDQN and dueling DQN are compatible with each other, they can be used together to form D3QN (utilizing a dueling architecture and double Q-learning) which can further improve stability and learning efficiency. Hence, in D3QN, the $Q(\mathcal{S}_t,\mathcal{A})$ can be calculated as:
\begin{equation}
Q(\mathcal{S}_t,\mathcal{A}_t) = V(\mathcal{S}_t) + A(\mathcal{S}_t,\mathcal{A}_t) - \frac{1}{|A|}\sum_{\mathcal{A}_{t+1}}A(\mathcal{S}_t,\mathcal{A}_{t+1})
\end{equation}
where $V(\mathcal{S}_t)$ is the state value of state $\mathcal{S}_t$, ${|A|}$ is the number of available actions in state $\mathcal{S}_t$, and $A(\mathcal{S}_t, \mathcal{A}_t)$ is the action advantage of taking action $\mathcal{A}_t$ in state $\mathcal{S}_t$, it can be calculate as follows:

\begin{equation}
A(\mathcal{S}_t,\mathcal{A}_t) = Q(\mathcal{S}_t,\mathcal{A}_t) - V(\mathcal{S}_t)
\end{equation}
 in which, if $A(\mathcal{S}_t,\mathcal{A}_t)$ is greater than 0, it indicates that taking action $\mathcal{A}_t$ can lead to better performance than taking the average action; otherwise, it means that taking action $\mathcal{A}_t$ is less effective than taking the average action. The update of D3QN is shown as follows:
\begin{equation}
\begin{aligned}
Q(\mathcal{S}_t,\mathcal{A}_t) \leftarrow Q(\mathcal{S}_t,\mathcal{A}_t) &+ \alpha \bigl[ \mathcal{R}_t + \gamma (\max_{\mathcal{A}}Q(\mathcal{S}_{t+1},\mathcal{A})) \\
&- Q(\mathcal{S}_t,\mathcal{A}_t)) \bigr]
\end{aligned}
\end{equation}
where $\alpha$ is the learning rate that is used to control the step size of updating neural network weights, and $\gamma$ is the discount factor that is used to measure the importance of future rewards. 

 Overall, compared to DQN, DDQN, and dueling DQN, D3QN can achieve more stable and efficient performance in a wide range of tasks, which makes D3QN suitable for real-world problems. Hence, we leverage D3QN as the model-free DRL model to propose our content importance evaluation model.

 \subsubsection{How the Evaluation Model Works}
 This subsection describes how the proposed content importance evaluation model works. The entire workflow involves four distinct parts:

 \begin{itemize}
  \item \textbf{Environment Observations}: We use software-defined network (SDN) technology to observe the network environment information from a global view. Each node's critical information, e.g., available link bandwidth, number of current user requests, popularity of each content, and modal type of contents, is periodically collected by the SDN controller. Therefore, the proposed evaluation model can retrieve all the needed information from the SDN controller which is implemented in a centralized server in the edge networks. Since we focus on the broader network dynamics and long-term trends rather than the instantaneous fluctuations in the network environment, the SDN controller does not need to collect critical information in real-time. Moreover, the collected information is statistical data which is much smaller than actual content data. Thus, the work load and overhead of the SDN controller is acceptable. We also utilize the SDN controller to manage the network.
  \item \textbf{Inputs and Outputs}: The inputs of the proposed content importance evaluation model include the available link bandwidth of a node, the number of user requests from that node, and a content's modal type. The output is the importance evaluation value of the content.
  \item \textbf{Trigger Mechanism}: There are two mechanisms to trigger the content importance evaluation model: 1) periodic mechanism; and 2) on-demand mechanism. The periodic mechanism is designed to fine-tune the evaluation results for minor changes in the network, aiming to maintain the high performance of the proposed caching mechanism. The trigger period is highly dependent on network dynamics. In this paper, as content popularity, user request patterns, etc., remain stable at the minute level, we can also set the trigger period at the minute level. Notably, the periodic mechanism is only used to fine-tune the evaluation results, which means it would not have a significant impact on the caching performance. Thus, the difference between a shorter trigger period and a longer trigger period would be insignificant. In the on-demand mechanism, the evaluation model will be triggered only if the network environment changes, such as the number of user requests changes dramatically in a short time period, there are newly released contents, etc. 
  \item \textbf{Content Importance Evaluation}: Once the content importance evaluation model can retrieve all the essential network environment information from the SDN controller, the trained D3QN agent can dynamically evaluate the importance of a given content based on the current network environment.
   \end{itemize}


\subsubsection{Implementation Details}
  
  This subsection outlines the implementation details of several techniques leveraged in the proposed content importance evaluation model to further enhance its stability and efficiency.

   \begin{itemize}
   \item \textbf{Exploration Technique:}
   How to find the trade-off between exploration and exploitation is a crucial aspect of training a DRL model. This is accomplished by utilizing the parameter $\epsilon$ to represent the probability that the agent will choose a random action, while selecting the action with the maximum Q-value with a probability of $1-\epsilon$. While more exploration can improve the chances of finding the optimal solution, it may also slow down convergence. Conversely, more exploitation can expedite convergence, but may lead to a sub-optimal solution. To obtain an optimal solution within an acceptable training time, it is essential to let the agent perform more exploration in the early training phase and perform more exploitation in the later phase. To achieve this, the agent is designed to follow the following strategy:
   \begin{equation}
   \epsilon = \begin{cases} 
      \epsilon_{start} & epi \leq E \\
      max(\epsilon \cdot dr, \epsilon_{end}) & epi > E \\
   \end{cases}
   \end{equation}
   where $\epsilon_{strat}$ is the maximum value of $\epsilon$, $\epsilon_{end}$ is the minimum value of $\epsilon$, $dr$ is the decay rate, $epi$ is the current episode, $E$ is a predefined threshold value. By adjusting $E$, we can decide the amount of explorations that the agent can perform.
   \item \textbf{Mapping Function:}
If content ID is treated as one of the elements in the state set, the proposed approach cannot be applied in dynamic networks due to the fact that newly released content is not covered in the state set of MDP, the previously trained model cannot guarantee its optimum for the new state set. To solve this issue, we remove the content ID from the state set. However, this leads the proposed content evaluation model to only return the importance values without content ID. Therefore, a mapping function is designed to solve this issue. The mapping function first removes the content ID from the SDN controller observed information to generate the inputs for the proposed content importance evaluation model, and stores the removed content ID temporally until the content importance is estimated. Once the content importance value is estimated, the mapping function will map the previously stored content ID with this importance value. In this way, the proposed importance evaluation model can estimate the newly released contents' importance.
   \item \textbf{Reward Function:}
   The reward function has a strong impact on the training process of the DRL model in terms of convergence and stability. In this article, we use the accumulated unsatisfied requests ratio to calculate the long-term rewards, and compare the immediate unsatisfied requests ratio with the predefined threshold to calculate the short-term rewards. An unsatisfied request is defined as a request where its QoS requirements are not satisfied by the network. By combing the long-term rewards and the short-term rewards, the agent can avoid getting stuck in a sub-optimal solution. By using multiple predefined thresholds to form a piecewise function, the rewards can be further distinguished, which can improve the speed of convergence and increase stability. The above calculations can be described as follows:
   \begin{equation}
   ratio = \frac{req_{uns}}{\sum_{epi=1}^{EPI} req_{epi}}
   \end{equation}
   \begin{equation}
       reward_{imm} = \begin{cases} 
      r_1 & ratio \leq th_1 \\
      r_2 & th_1 < ratio  \leq th_2 \\
      -r_2 & th_2 < ratio  \leq th_3 \\
      -r_1 & ratio \geq th_3 \\
   \end{cases}
   \end{equation}
   \begin{equation}
   reward_{acc} = reward_{acc}^\prime + reward_{imm}
   \end{equation}
   where $req_{uns}$ is the number of unsatisfied requests, $req_{epi}$ is the total number of requests in episode $epi$, $EPI$ is the maximum number of training episodes, $reward_{imm}$ stands for the immediate rewards after the agent choose an action, $reward_{acc}$ represents the accumulated rewards, $reward_{acc}^\prime$ represents the previous accumulated rewards before taking the current action, $r_1$ and $r_2$ are configurable integer values, $th_1, th_2$, and $th_3$ are predefined thresholds. To further improve the speed of convergence and increase stability, those thresholds can be defined based on the performance of existing caching approaches.
   \item \textbf{Deployment Details:}
   Since training a DRL model needs significant resources and time, we train the proposed importance evaluation model offline. Once the model has been trained, it can be deployed in the network. Considering that network edge devices, (e.g., BSs) are usually resource-constrained, the trained model is deployed in the centralized server which also holds the SDN controller. Consequently, the trained model can easily fetch the current network environment information from the SDN controller, and distribute the evaluated importance value of each content to all network edge devices. 
   \end{itemize}

 \subsection{Content Importance-based Caching Model}
 In this subsection, we present the details of the proposed content importance-based caching model which consists of a content importance-based caching decision policy and a content importance-based replacement policy.

\subsubsection{Caching Decision Policy}
The goal of the caching decision policy is to decide what content should be cached locally. This decision can be affected by several factors such as the content's popularity, the content's size, available link bandwidth, the content's modal type, etc. To address this issue, we leverage the proposed content importance evaluation model that takes these factors into account to provide a comprehensive evaluation of the content's importance. 

Based on this evaluated content importance, we propose a content importance-based caching decision policy that can cache content with high importance value. More specifically, the caching node needs to maintain a key-value table for storing its cached contents' importance value and size information for making quick caching decisions. If a newly arrived content's size is less than the available caching space (denoted as $ac_s(e_n), ac_s(e_n) \leq c_{size}(en)$, where $c_{size}(e_n)$ is the maximum caching space of node $e_n$), then the content can be directly cached at this caching node, and its importance value and size information will be added to the key-value table. Otherwise, the caching node needs to check if the content's importance value is greater than the minimum importance value of the cached content. If yes, then the newly arrived content can be cached locally; otherwise, the caching node will not cache it. Notably, comparing the newly arrived content's importance value with cached contents' importance value means there is not enough space for caching the newly arrived content. Thus, a replacement policy is required to evict less important contents to free up enough caching space to cache the newly arrived content.

\subsubsection{Replacement Policy}
A content importance-based replacement policy is designed to cooperate with the proposed caching decision policy. Once the available caching memory is insufficient to cache the newly arrived important content, the replacement policy will be triggered. Firstly, the caching node needs to sort all cached contents based on their importance value. Then, the caching node can evict contents based on sorted importance value until there is enough caching space for caching the newly arrived content. Meanwhile, the caching node needs to remove the evicted contents' information from the key-value table.

The algorithm of the proposed caching model can be found in Algorithm \ref{model} which includes both the caching decision policy and the replacement policy.




\begin{algorithm}[htbp]
\setcounter{AlgoLine}{0}
\LinesNumbered
\SetSideCommentRight
\SetNoFillComment
\Indm\Indm
\Indp\Indp

\vspace{1mm}
\textbf{Input: }{The incoming content $d_m=[o(d_m), r_{bw}(d_m), r_l(d_m), s(d_m)]$, current network state $\mathcal{S}_t = \{a(e_nt, t), req_{d_m}(e_n, t), Req(b_n, t), o(d_m), s(d_m)\}$, caching node's caching table $T_c$ and its available caching space $ac_s(e_n)$}\\
\textbf{Output: }{Caching decision}
\vspace{1mm} \\
\hrule
\vspace{1mm}
Content importance evaluation model evaluates the incoming content $d_m$'s importance $I(d_m, e_n, t)$ based on $\mathcal{S}_t$\;
Sort table $T_c$ in increasing order of cached contents' importance value\;
\eIf{$s \leq ac_s$ }
{cache this content directly\;}
{
    $item \gets 1$\;
    \While{$s > ac_s$}{
        \eIf{importance value of $T_c[item] < I(d_m, e_n, t)$}{
            evict $T_c[item]$ from local caching space\;
            $ac_s \gets ac_s +$ content size of $T_c[item]$\;
            \If{item $<$ number of items in $T_c$}{$item \gets item + 1$\;}
        }{
            break\;
        }
    }
    \eIf{$s \leq ac_s$}{
        cache this content\;
    }{
        do not cache this content\;
    }
}
\vspace{2.5mm}
\caption{Caching Decision and Replacement Policies of the Proposed Caching Model}
\label{model}
\end{algorithm}

\section{Performance Evaluation}
In this section, we first describe our simulation settings. Then we introduce the evaluation metrics and we finally present and discuss the simulation results.  

\subsection{Simulation Settings}
\begin{figure}[!htp]
\centering{\includegraphics[width=0.5\textwidth]{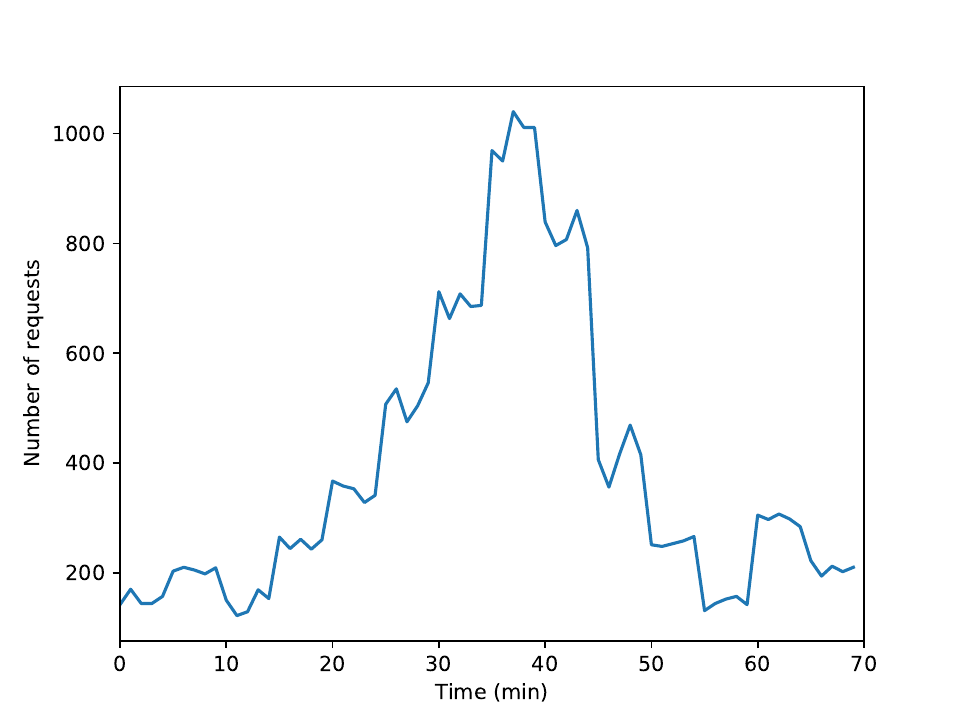}}
\caption{Pattern of user requests} 
\label{request} 
\end{figure}

We use the network depicted in Fig. \ref{network-model} for our simulation. The backbone link bandwidth $bl_{bw}$ between edge nodes and the backbone network is set as 100 Gbps. Three types of multi-modal contents are used in our simulation: video contents, audio contents, and haptic contents. Their bandwidth and latency requirements are listed in Table \ref{QoS}. To mimic a dynamic network environment which includes idle periods and peak periods, we vary the number of user requests over time. We generate the user requests based on \cite{iCache}, which is shown in Fig. \ref{request}. Additionally, as time progresses, we also randomly generate new contents and associated requests to better emulate a dynamic network environment. There are 500 contents in total, and users have requested them 24,412 times. This dataset size aligns with typical practices in existing DRL-based approaches, such as \cite{reviewer-3-1} and \cite{reviewer-3-2}.


\begin{table}[h]
    \centering
    \caption{QoS Requirements for Multi-modal Contents}
    \label{QoS}
    \renewcommand{\arraystretch}{1.5}
    \begin{tabular}{|c|c|c|}
        \hline
        \textbf{Content Type} & \textbf{Bandwidth} & \textbf{Latency} \\
        \hline
        1080P 30 fps video & 8-15 Mbps & $<$100 ms \\
        \hline
        1080P 60 fps video & 12-24 Mbps & $<$100 ms \\
        \hline
        4K 30 fps video & 25-50 Mbps & $<$100 ms \\
        \hline
        4K 60 fps video & 50-100 Mbps & $<$100 ms \\
        \hline
        8K 30 fps video & 100-150 Mbps & $<$100 ms \\
        \hline
        8K 60 fps video & 150-200 Mbps & $<$100 ms \\
        \hline
        MPEG-1 Audio Layer III (MP3) audio & 64-320 kbps & $<$100 ms \\
        \hline
        Blue-ray quality audio & 448 kbps & $<$10 ms \\
        \hline
        Home theater quality audio & 1-6 Mbps & $<$50 ms \\
        \hline
        Low-fidelity haptic & <100 kbps & $<$10 ms \\
        \hline
        High-fidelity haptic & $<$1 Mbps & $<$1 ms \\
        \hline
    \end{tabular}
\end{table}

For comparisons, we choose LCE (leave copy everywhere) \cite{LCE}, DPWCS (dynamic popularity window-based caching scheme) \cite{DPWCS}, and CP-DQN \cite{semantic} as the baseline caching decision policies and choose LRU (least recently used) as the replacement policy for them. A federated DRL-based replacement policy is proposed in \cite{DDQN}. Since the federated technique has relatively minor effect on caching performance, we use DDQN to represent the proposed approach in \cite{DDQN}, and also choose it as a baseline approach. We also remove the content modal feature from the state set of the proposed approach to evaluate the impact of content modality (denoted as no-Modality). In LCE, caching nodes will cache every content that goes through them. DPWCS employs a configurable popularity window to track the content popularity and dynamically adjusts the caching threshold to cache popular contents, resulting in superior efficiency compared to LCE. CP-DQN is a typical DQN-based caching approach that defines which content should be cached at which node as the action set. 

 We perform the simulation on a Linux GPU server equipped with an Intel(R) Xeon(R) Platinum 8352V CPU (12 cores, 2.1 GHz), an RTX 4090 (24GB) GPU, and 90 GB of memory, using Python 3.9 and PyTorch 2.0. Since the DRL model cannot obtain a fixed result for comparisons, we perform 30 different runs for the proposed caching approach and calculate the average over these 30 runs. For simplicity, we assume that contents belonging to the same modality are partitioned into packets of the same size for transmission. The details of the parameter settings are listed in Table \ref{parameters}.

 \begin{table}[h]
    \centering
    \caption{Parameter Settings}
    \label{parameters}
    \renewcommand{\arraystretch}{1.5}
    \begin{tabular}{|c|c|}
        \hline
        \textbf{Description} & \textbf{Value} \\
        \hline
        Maximum number of episodes ($EPI$) & 3000   \\
        \hline
        Value of starting $\epsilon$ ($\epsilon_{start}$) &  0.99\\
        \hline
        Value of ending $\epsilon$ ($\epsilon_{end}$) &  0.01 \\
        \hline
        Decay rate ($dr$) & 0.997  \\
        \hline
        Number of edge caching nodes ($N$) & 6  \\
        \hline
        Access link bandwidth ($al_{bw}$) & 1 Gbps \\
        \hline
        Backbone link bandwidth ($bl_{bw}$) & 100 Gbps \\
        \hline
        Learning rate ($\alpha$) & 0.0005  \\
        \hline
        Discount factor ($\gamma$) & 0.99 \\
        \hline
        Duration time of user requests ($T$) & 8 mins \\
        \hline
    \end{tabular}
\end{table}


\subsection{Evaluation Metrics}
We adopt the \emph{average number of hops}, \emph{hit ratio}, \emph{reduced load ratio}, and the \emph{unsatisfied requests ratio} as the evaluation metrics to evaluate the performance of the proposed DRL-based caching scheme. 

\subsubsection{Average Number of Hops} is used in this article to measure the content retrieval latency since high efficiency caching schemes can cache important contents at the network edge to reduce content retrieval latency. The average number of hops can be calculated as follows:
   \begin{equation}
avgHop = \sum_{e_n \in E}\sum_{m \in M}\sum_{t \in T}\frac{hop[req_{d_m}(e_n, t)]}{req_{d_m}(e_n, t)}
   \end{equation}
where $hop[req_{d_m}(e_n, t)]$ is number of hops for requests $req_{d_m}(e_n, t)$. As defined in Section \uppercase\expandafter{\romannumeral3}, $req_{d_m}(e_n, t)$ is the total number of requests for multi-modal contents from edge node $e_n$ at time $t$.

\subsubsection{Hit Ratio} is the most widely used metric to evaluate the efficiency of caching schemes as it reflects the proportion of requests that the cache can fulfill. It can be calculated by:
   \begin{equation}
hitRatio = \sum_{e_n \in E}\sum_{m \in M}\sum_{t \in T}\frac{cache[req_{d_m}(e_n, t)]}{req_{d_m}(e_n, t)}
   \end{equation}
where $cache[req_{d_m}(e_n, t)]$ is the number of requests $req_{d_m}(e_n, t)$ that can be satisfied by caching.

\subsubsection{Reduced Load Ratio} indicates the proportion of traffic loads that are reduced by leveraging caching. Compared with the hit ratio, the reduced load ratio considers the size of cached contents. Hence, it can directly reflect how the caching scheme can alleviate the network traffic load. It can be calculated by:
   \begin{equation}
rLRatio = \sum_{e_n \in E}\sum_{m \in M}\sum_{t \in T}\frac{z_{d_m} \cdot cache[req_{d_m}(e_n, t)]}{req_{d_m}(e_n, t)}
   \end{equation}
   where $z_{d_m} \cdot cache([req_{d_m}(e_n, t))]$ stands for the total size of cached contents that can serve requests $req_{d_m}(e_n, t)$.

\subsubsection{Unsatisfied Requests Ratio} represents the proportion of contents whose QoS requirements are not satisfied. This metric can be utilized to assess the efficiency of caching schemes since caching increases the likelihood of satisfying the QoS requirements of the contents. Additionally, for multi-modal contents, caching one type of content can affect other types. For instance, caching videos can reduce network bandwidth usage but may not meet the ultra-low latency requirements for haptic contents due to long-distance transmission. Conversely, caching haptic contents can solve this issue, but it may not satisfy the high bandwidth requirements for video contents, as transmitting numerous videos consumes a significant amount of bandwidth. The unsatisfied requests ratio can be calculated as follows:
   \begin{equation}
unSatRatio = \sum_{e_n \in E}\sum_{m \in M}\sum_{t \in T}\frac{unSat[req_{d_m}(e_n, t)]}{req_{d_m}(e_n, t)}
   \end{equation}
where $unSat[req_{d_m}(e_n, t)]$ is the number of unsatisfied requests in $req_{d_m}(e_n, t)$.

\subsection{Evaluation Results}
We compare the efficiency of our proposed content importance-based caching scheme with five other caching schemes: LCE, DDQN, DPWCS, CP-DQN, and our proposed approach without considering the content modal feature (denoted as no-Modality). We evaluate their performance in terms of the average number of hops, hit ratio, reduced load ratio, and unsatisfied requests ratio.

\begin{figure}[!htp]
\centering{\includegraphics[width=0.4\textwidth]{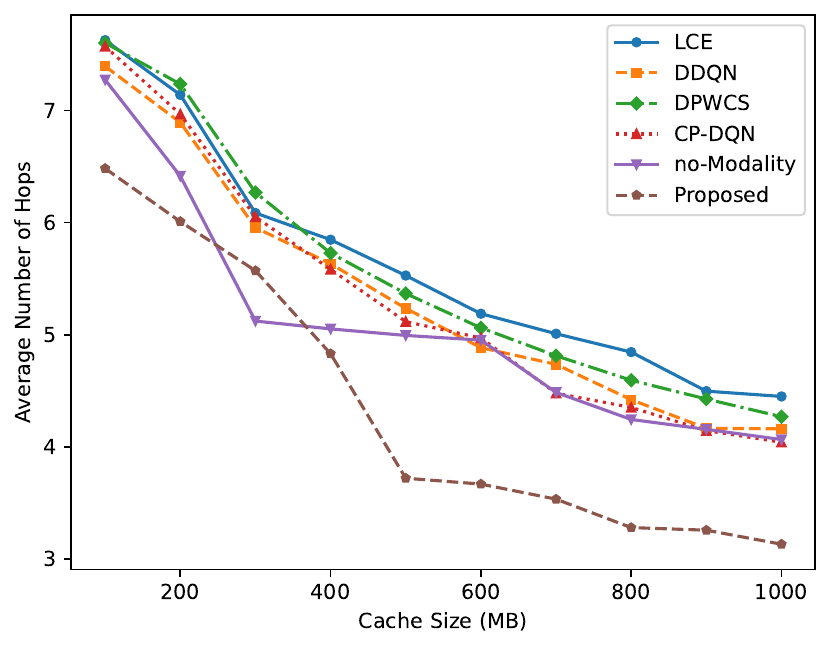}}
\caption{Impact of cache size of edge nodes on average number of hops} 
\label{latency} 
\end{figure}
Fig. \ref{latency} illustrates how the performance of these six caching schemes changes in terms of the average number of hops with different cache size settings. Notably, the proposed content importance-based caching scheme outperforms baseline approaches in terms of the average number of hops. For example, when the edge node's cache size is set to 600 MB, the proposed caching scheme can reduce the average number of hops by 27\% and 25\% compared with CP-DQN and no-Modality respectively. Although CP-DQN and no-Modality have similar performance overall, no-Modality achieves better performance with small cache sizes (<600 MB) in terms of the average number of hops. Since CP-DQN and DDQN are modeled from an optimization perspective, which makes them less adaptable to network environment changes; while no-Modality does not consider the content's modality, which makes it cannot detect the impact of the content's modality. Consequently, the effectiveness of all the DRL-based approaches is degraded. Additionally, the performance of LCE is the worst due to the fact that it cannot filter the popular contents to cache.


\begin{figure}[!htp]
\centering{\includegraphics[width=0.4\textwidth]{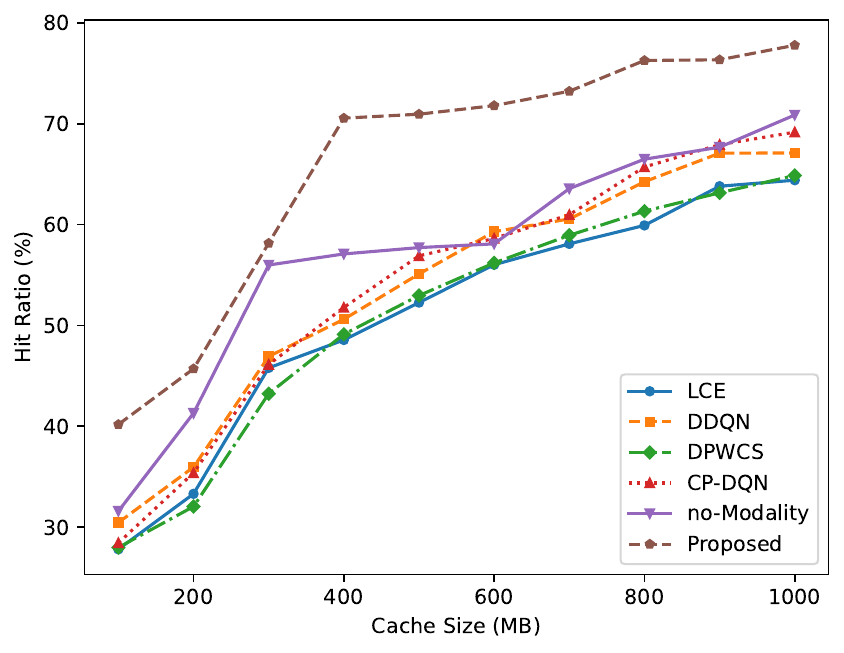}}
\caption{Impact of cache size of edge nodes on hit ratio} 
\label{hit} 
\end{figure}

Fig. \ref{hit} demonstrates how cache size affects the performance of these six caching schemes in terms of hit ratio. Obviously, all caching schemes can achieve a higher hit ratio with the increase of cache size. Obviously, our proposed approach can significantly improve the caching performance in terms of hit ratio. Especially, when the edge node's cache size is set to 400 MB, the proposed caching scheme can achieve at least a 15\% higher hit ratio than baseline approaches. The reason why the proposed approach can outperform other baseline approaches is that it can recognize vital haptic contents to cache. Since haptic contents are usually small in size, the caching space utilization rate will be significantly improved. Additionally, the performance of no-Modality ranks as the 2nd place. The reason for this change is that the reward function of no-Modality has been revised to maximize the hit ratio, given the removal of content modality from the state set. Consequently, the original reward function of the proposed approach is no longer applicable to no-Modality. Moreover, although CP-DQN cannot maintain effectiveness in dynamic networks, it can still slightly outperform conventional approaches, such as LCE, DDQN, and DPWCS.


\begin{figure}[!htp]
\centering {\includegraphics[width=0.4\textwidth]{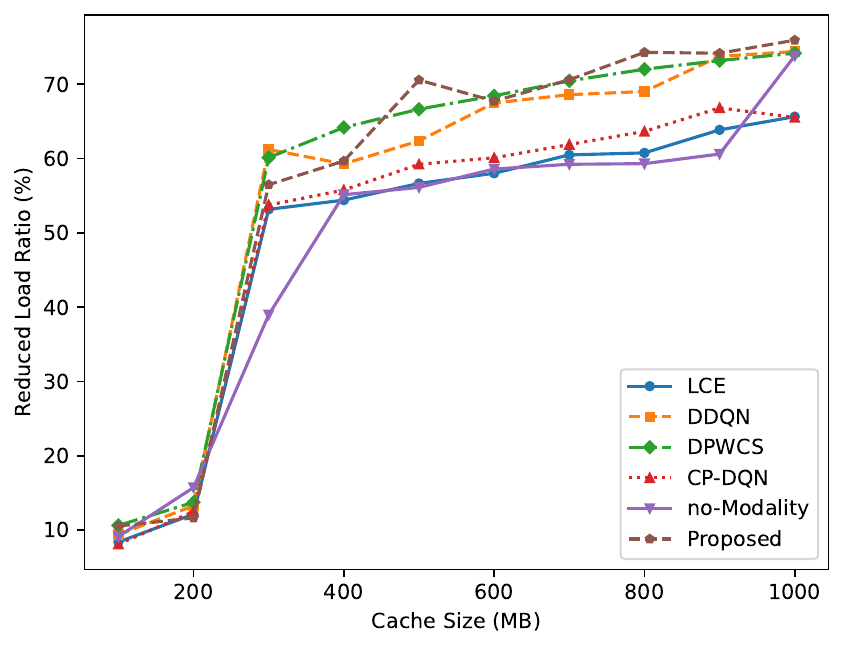}}
\caption{Impact of cache size of edge nodes on reduced network load} 
\label{load} 
\end{figure}	
 The impact of the cache size of edge nodes on the reduced network load is presented in Fig. \ref{load}. We can see that there is not too much difference among all these six approaches when the cache size is less than 300 MB. When cache size is greater than 300 MB, the performance of these six approaches can be classified into two groups. The first group includes DDQN, DPWCS, and the proposed caching scheme. All of them can obtain a better reduced load ratio compared to the other group which includes LEC, CP-DQN, and no-Modality. However, it is worth noting that when the cache size is greater than 900 MB, the performance of no-Modality transitions into the first group, demonstrating comparable performance to the approaches in the first group. Overall, our proposed approach has a very close performance with DPWCS. The reason behind this phenomenon is that the objective of our proposed approach is to minimize the unsatisfied requests ratio rather than minimizing the reduced network load. Hence, it does not demonstrate superiority in terms of the reduced network load ratio.

\begin{figure}[!htp]
\centering{\includegraphics[width=0.4\textwidth]{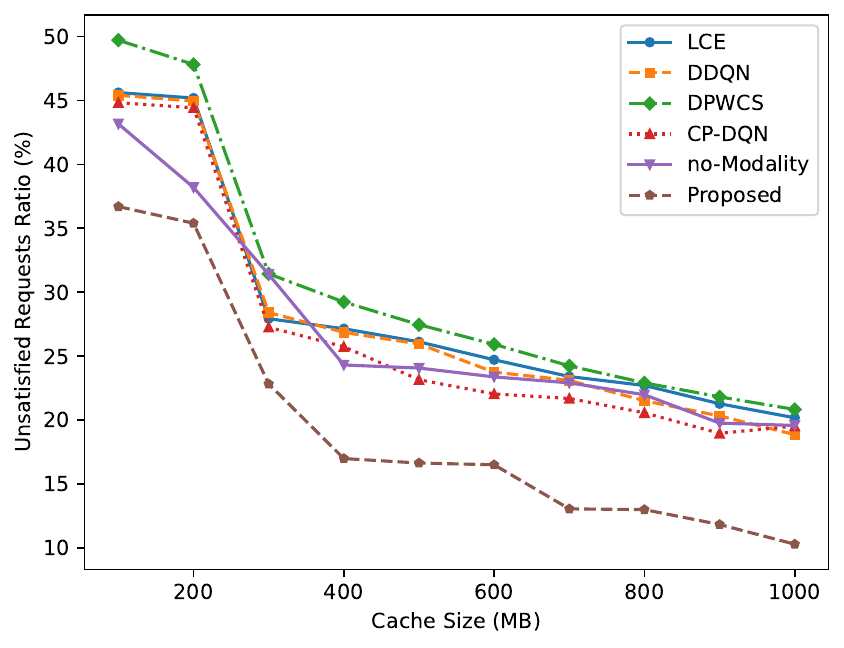}}
\caption{Impact of cache size of edge nodes on unsatisfied requests ratio} 
\label{unsat} 
\end{figure}

From Fig. \ref{unsat} we can see that our proposed caching scheme can always achieve the best performance in terms of the unsatisfied requests ratio among all caching schemes. The reason is that the objective of the proposed caching scheme is to minimize the unsatisfied requests ratio to enhance users' QoE. Notably, the performance gap between the proposed scheme and no-Modality widens as the edge node's cache size increases beyond 200 MB. In particular, when the cache size is configured at 1000 MB, our proposed caching scheme exhibits remarkable superiority. It can achieve a reduction of approximately 47\% in the unsatisfied requests ratio compared to CP-DQN and no-Modality. Furthermore, it outperforms non-AI-based approaches like DPWCS by more than twice the margin. The closely matched performance of the other three DRL-based approaches can be attributed to their neglect of the influence of content modality. These results unequivocally illustrate the effectiveness of our proposed caching scheme, particularly when the edge node possesses a large cache size.

\begin{figure*}[htbp]
  \centering
  \subfigure[Comparisons among approaches in terms of average number of hops]{
    \label{1}
    \includegraphics[width=0.48\textwidth]{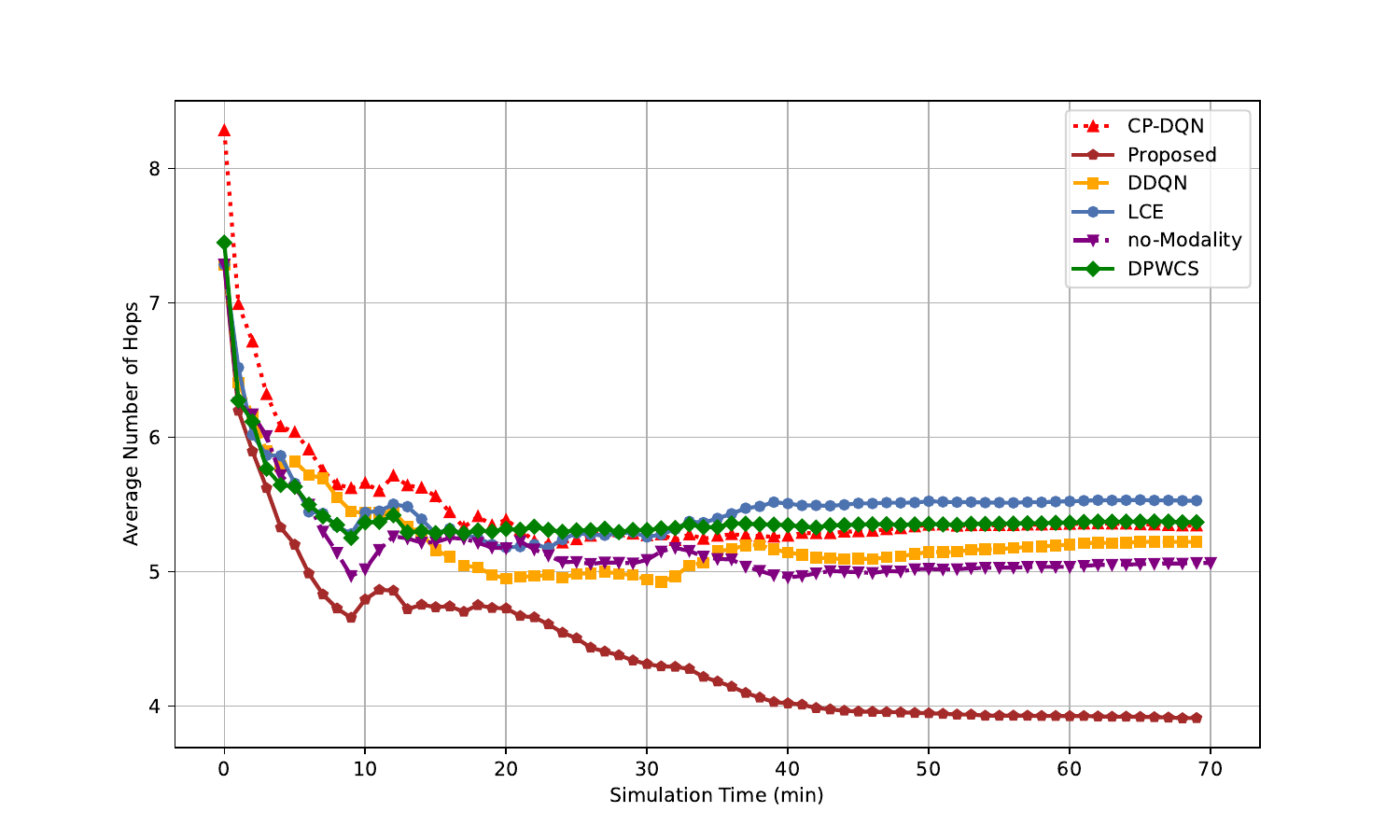}
  }
  \subfigure[Comparisons among approaches in terms of hit ratio]{
    \label{2}
    \includegraphics[width=0.48\textwidth]{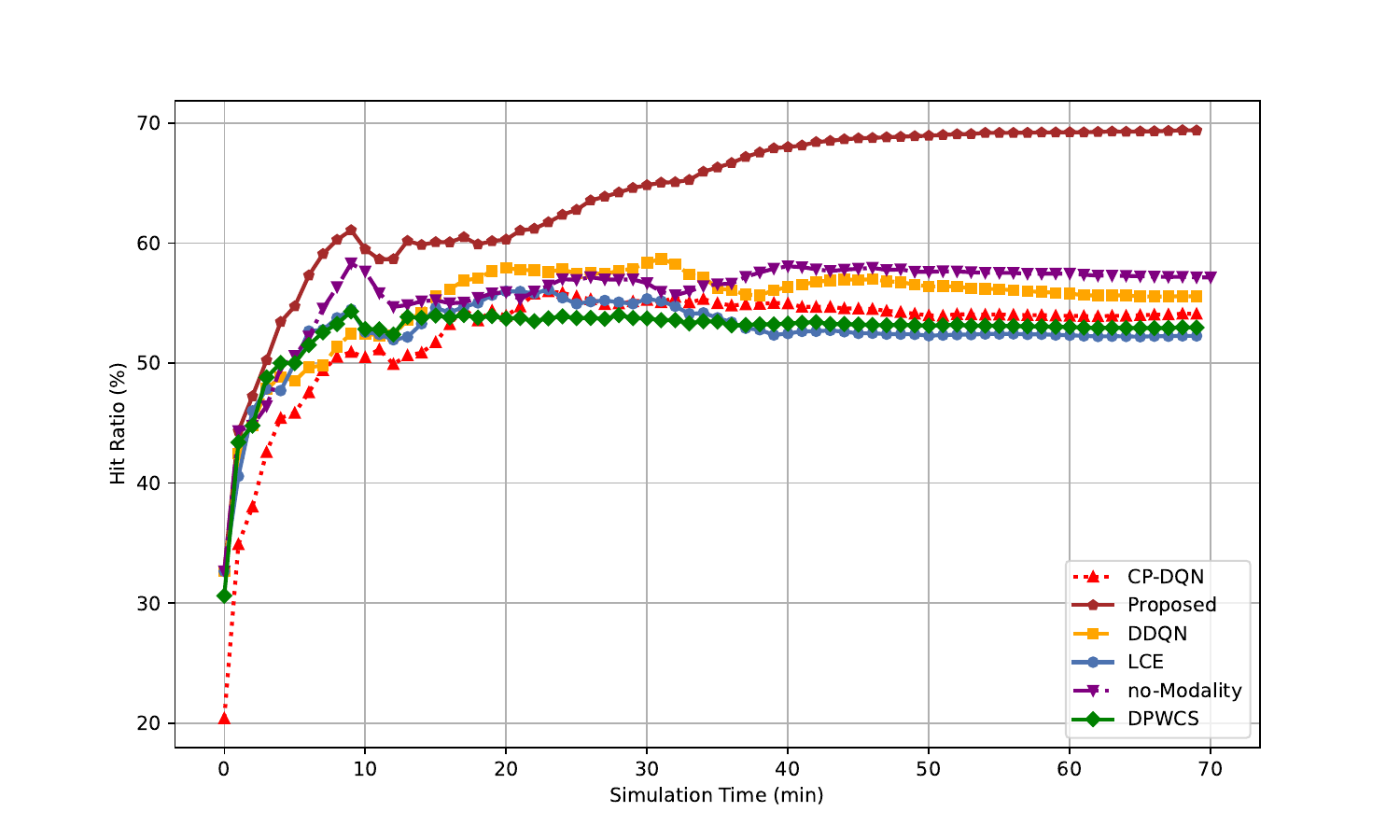}
  }
  \subfigure[Comparisons among approaches in terms of reduced load ratio]{
    \label{3}
    \includegraphics[width=0.48\textwidth]{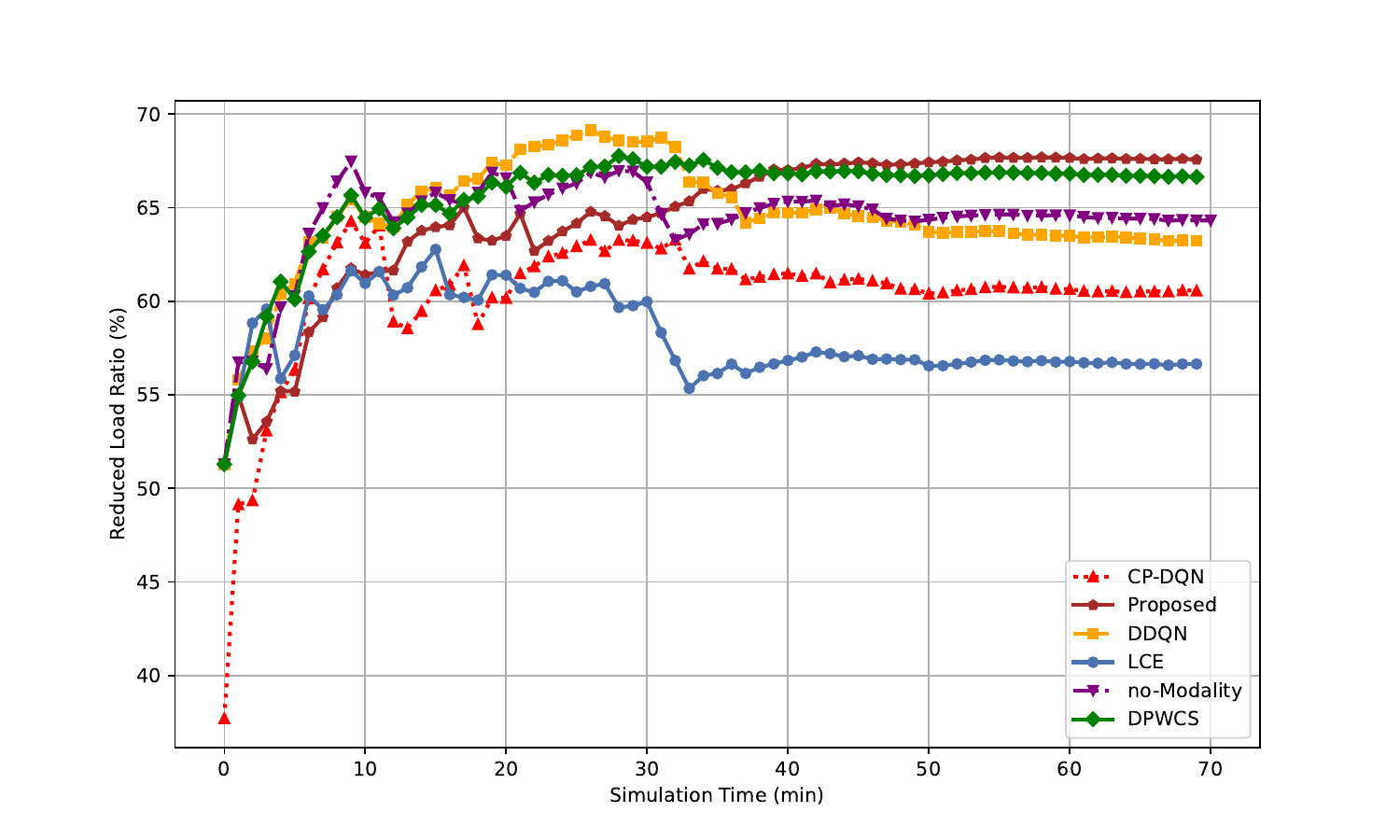}
  }
  \subfigure[Comparisons among approaches in terms of unsatisfied requests ratio]{
    \label{4}
    \includegraphics[width=0.48\textwidth]{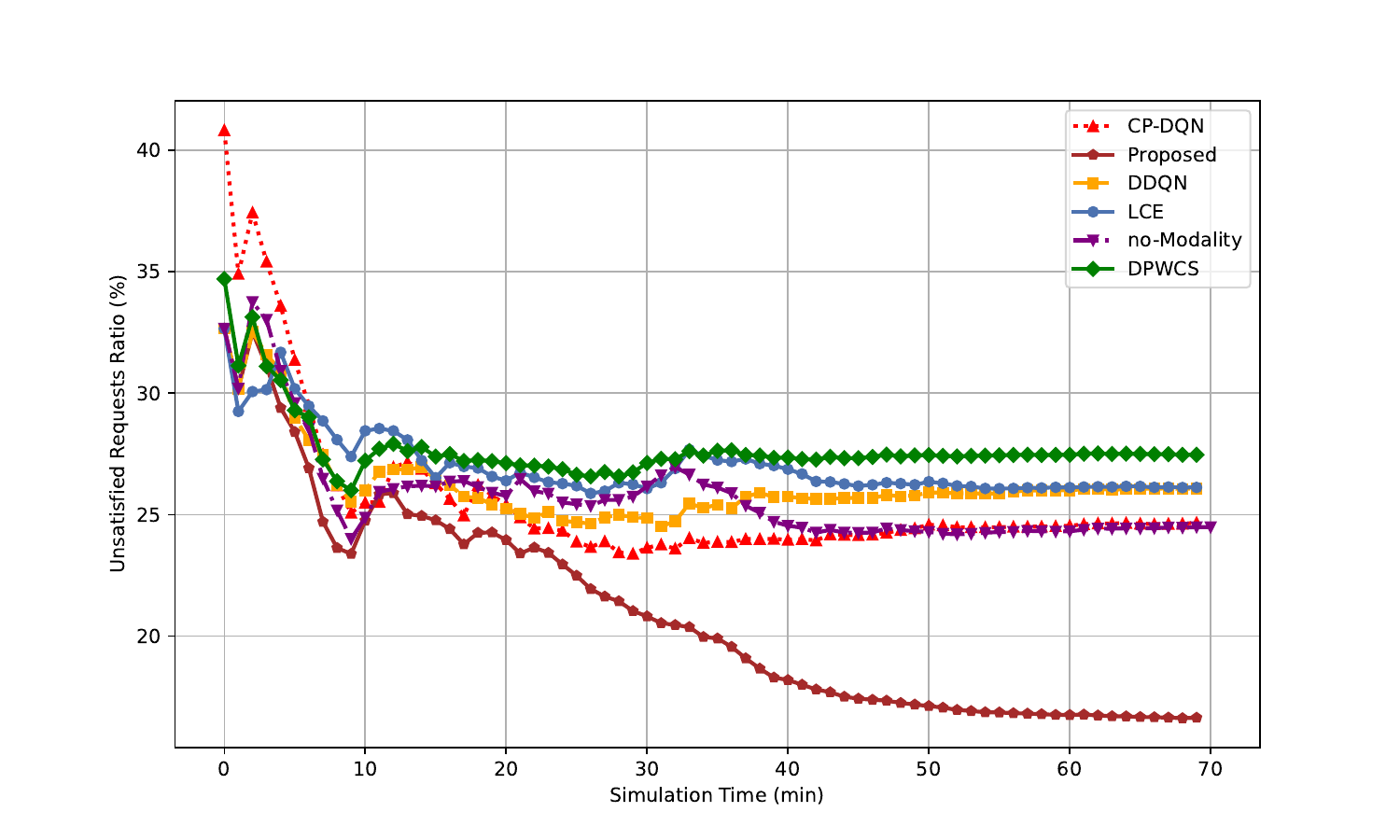}
  }
  \caption{Performance evolution of all approaches over the simulation time}
  \label{overtime}
  \end{figure*}

Fig. \ref{overtime} illustrates the performance of these six approaches over time in terms of the average number of hops, hit ratio, reduced load ratio, and unsatisfied requests ratio. Obviously, the proposed approach obtains the best performance during the entire simulation time in terms of all the simulation metrics. Fig. \ref{overtime}.(c) indicates that the proposed approach also demonstrates slightly superiority in terms of reduced load ratio in the late stages of the simulation (e.g., after 40 mins), while showing less effectiveness in the initial phase of the simulation (e.g., between 10 and 40 mins) compared to DDQN and DPWCS. This is primarily due to the proposed approach focuses on minimizing unsatisfied requests ratio rather than reduced load ratio.

All sub-figures in Fig. \ref{overtime} show a notable fluctuation in the performance of all approaches occurs around 10 min. This fluctuation is due to the first change in user request patterns (as shown in Fig. \ref{request}), and all approaches are reactive caching approach, i.e., they only respond to user requests rather than making decisions before user requests arrival. However, as simulation progresses (e.g., between 10 and 30 mins), the proposed approach demonstrates a consistent improvement trend, while other approaches continue to show fluctuations. As the simulation advances to the 30-40 min, a drastic change in user request patterns causes a significant degradation in the performance of baseline approaches. In contrast, the proposed approach continues to steadily enhance its performance until achieving a stable performance after 50 mins. Notably, other approaches cannot improve their performance after 40 min.

\section{Conclusion}
In this article, we proposed a content importance-based caching scheme which can automatically choose the most important contents for multi-modal services in dynamic network environments. Although existing DRL-based caching schemes, such as \cite{DRL-1, DRL-2, DRL-3}, can obtain optimal caching decisions, they cannot guarantee their optimum for dynamic network environments that involve newly generated contents, varying user request patterns, varying available link bandwidth, etc. To overcome this issue, we proposed a content importance evaluation model to dynamically evaluate contents' importance in dynamic network environments by leveraging D3QN model. Based on the evaluated importance, we can easily make caching decisions and replacement decisions for multi-modal contents in dynamic network environments.

The evaluation results illustrated that our proposed approach can achieve better performance in terms of the average number of hops, hit ratio, reduced load ratio, and unsatisfied requests ratio compared with baseline approaches (LCE, DDQN, CP-DQN, and DPWCS). 

For future work, one possible direction is to consider the correlations among three modalities of contents to design caching schemes. For example, recent AI technologies can generate one modal content from other modal contents, which means that the content's importance will be affected by more factors. Therefore, considering the correlations among different modalities of contents to design more efficient caching schemes could be a potential research direction.


\bibliographystyle{IEEEtran}
\bibliography{references}


%





\ifCLASSOPTIONcaptionsoff
  \newpage
\fi

\begin{IEEEbiography}
[{\includegraphics[width=1in,height=1.25in,clip,keepaspectratio]{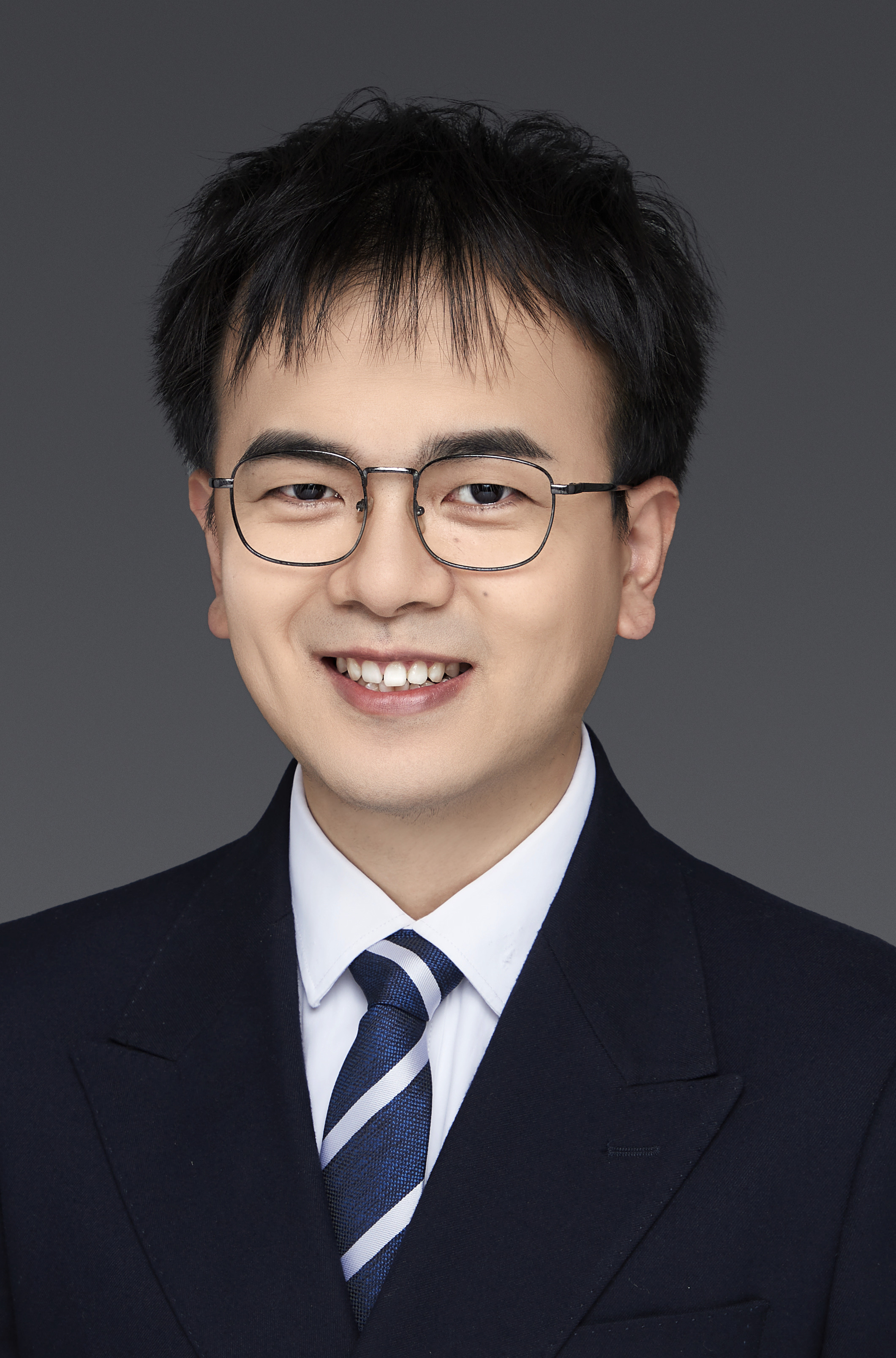}}]{Zhe Zhang}
(Member, IEEE) received his Ph.D. degree in Electrical and Computer Engineering from Carleton University, Ottawa, Canada in 2019. Currently, he is an Assistant Professor with the School of Communications and Information Engineering, Nanjing University of Posts and Telecommunications (NJUPT), Nanjing, Jiangsu, China. Before he joined NJUPT, he was a research engineer at Huawei Ottawa Research Center, Ottawa, Canada. His research interests include multimedia networking, intelligent networks, in-network caching, software-defined networking, and Internet of things.
\end{IEEEbiography}
 \vspace{-0.2cm}
\begin{IEEEbiography}
[{\includegraphics[width=1in,height=1.25in,clip,keepaspectratio]{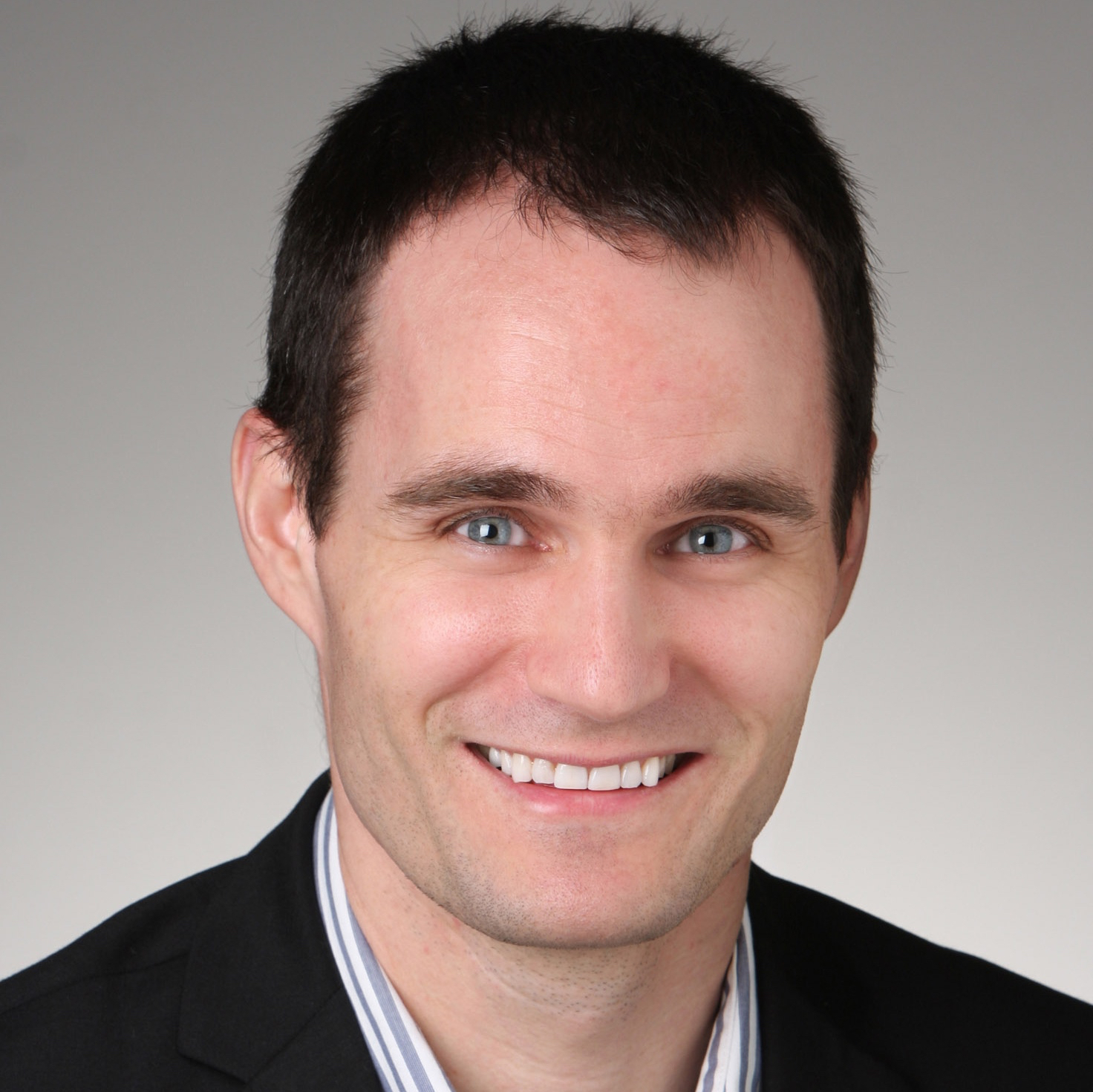}}]{Marc St-Hilaire}
(Senior Member, IEEE) received the Ph.D. degree in computer engineering from Polytechnique Montreal in 2006. He is currently a Professor with the School of Information Technology with a cross-appointment to the Department of Systems and Computer Engineering at Carleton University, Ottawa, Canada. He is conducting research on various aspects of wired and wireless communication systems. More precisely, he is interested in network planning and design, network architecture, mobile computing, and cloud/edge computing. With more than 170 publications, his work has been published/presented in several journals and international conferences. Over the years, he has received several awards, including the Carleton Faculty graduate mentoring award, the Carleton teaching achievement award, and several best paper awards. Finally, Dr. St-Hilaire is actively involved in the research community. In addition to serving as a member of technical program committees of various conferences, he is equally involved in the organization of several national and international conferences and workshops. He is also a senior member of the IEEE.
\end{IEEEbiography}
 \vspace{-2cm}
\begin{IEEEbiography}
[{\includegraphics[width=1in,height=1.25in,clip,keepaspectratio]{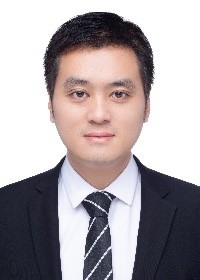}}]{Xin Wei}
(Member, IEEE) received his Ph.D. degree major at information and communication engineering from Southeast University, Nanjing, China in 2009. Now, he is a professor in Nanjing University of Posts and Telecommunications, China. His research interests are in the areas of multimedia communications and computing, educational technology.
\end{IEEEbiography}
 \vspace{-2cm}
 
\begin{IEEEbiography}
[{\includegraphics[width=1in,height=1.25in,clip,keepaspectratio]{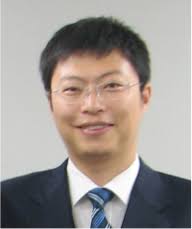}}]{Haiwei Dong}
(Senior Member, IEEE) received the Ph.D. degree from Kobe University, Kobe, Japan in 2010 and the M.Eng. degree from Shanghai Jiao Tong University, Shanghai, China, in 2008. He was a Principal Engineer with Artificial Intelligence Competency Center in Huawei Technologies Canada, Toronto, ON, Canada, a Research Scientist with the University of Ottawa, Ottawa, ON, Canada, a Postdoctoral Fellow with New York University, New York City, NY, USA, a Research Associate with the University of Toronto, Toronto, ON, Canada, and a Research Fellow (PD) with the Japan Society for the Promotion of Science, Tokyo, Japan. He is currently a Principal Researcher with Ottawa Research Center, Huawei Technologies Canada, Ottawa, ON, Canada, and a registered Professional Engineer in Ontario. His research interests include artificial intelligence, multimedia, metaverse, and robotics. He also serves as a Column Editor of IEEE Multimedia Magazine; an Associate Editor of ACM Transactions on Multimedia Computing, Communications, and Applications; and an Associate Editor of IEEE Consumer Electronics Magazine.
\end{IEEEbiography}
 \vspace{-2cm}
 
\begin{IEEEbiography}
[{\includegraphics[width=1in,height=1.25in,clip,keepaspectratio]{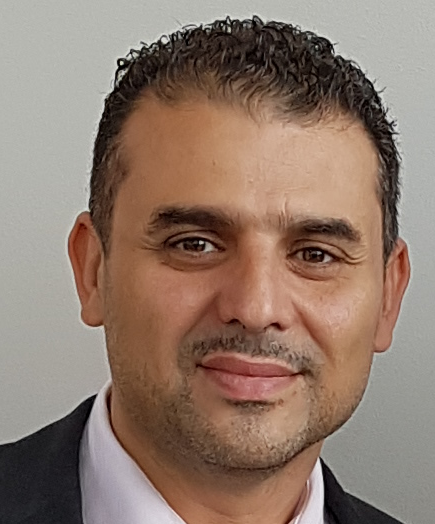}}]{Abdulmotaleb El Saddik}
(Fellow, IEEE) is currently a Distinguished Professor with the School of Electrical Engineering and Computer Science, University of Ottawa and a Professor with Mohamed bin Zayed University of Artificial Intelligence. He has supervised more than 120 researchers. He has coauthored ten books and more than 550 publications and chaired more than 50 conferences and workshops. His research interests include the establishment of digital twins to facilitate the well-being of citizens using AI, the IoT, AR/VR, and 5G to allow people to interact in real time with one another as well as with their smart digital representations. He received research grants and contracts totaling more than \$20 M. He is a Fellow of Royal Society of Canada, a Fellow of IEEE, an ACM Distinguished Scientist and a Fellow of the Engineering Institute of Canada and the Canadian Academy of Engineers. He received several international awards, such as the IEEE I\&M Technical Achievement Award, the IEEE Canada C.C. Gotlieb (Computer) Medal, and the A.G.L. McNaughton Gold Medal for important contributions to the field of computer engineering and science.
\end{IEEEbiography}
 \vspace{-2cm}



\end{document}